# Eshelby tensors and overall properties of nano-composites considering both interface stretching and bending effects

Junbo Wang[1], Peng Yan[1], Leiting Dong[1*], Satya N. Atluri [2]

[1]School of Aeronautic Science and Engineering, Beihang University, Beijing 100191, CHINA

[2] Department of Mechanical Engineering, Texas Tech University, USA

**Abstract:** In this study, the fundamental framework of analytical micromechanics is generalized to consider nano-composites with both interface stretching and bending effects. The interior and exterior Eshelby tensors for a spherical nano-inclusion, with an interface defined by the Steigmann–Ogden model, subjected to an arbitrary uniform eigenstrain are derived for the first time. Correspondingly, the stress/strain concentration tensors for a spherical nano-inhomogeneity subjected to arbitrary uniform far-field stress/strain loadings are also derived. Using the obtained concentration tensors, the effective bulk and shear moduli are derived by employing the dilute approximation and the Mori-Tanaka method, respectively, which can be used for both nano-composites and nano-porous materials. An equivalent interface curvature parameter reflecting the influence of the interface bending resistance is found, which can significantly simplify the complex expressions of the effective properties. In addition to size-dependency, the closed form expressions show that the effective bulk modulus is invariant to interface bending resistance parameters, in contrast to the effective shear modulus. We also put forward for the first time a characteristic interface curvature parameter, near which the effective shear modulus is affected significantly. Numerical results show that the effective shear moduli of nano-composites and nano-porous materials can be greatly improved by an appropriate surface modification. Finally, the derived effective modulus with the Steigmann–Ogden interface model is provided in the supplemental MATLAB code, which can be easily executed, and used as a benchmark for semi-analytical solutions and numerical solutions in future studies.

[*] Corresponding author: ltdong@buaa.edu.cn (L. Dong). Address: School of Aeronautic Science and Engineering, Beihang University, Beijing, 100191, CHINA.

1. **Introduction**

In recent decades, the interest in modelling surfaces and interfaces is growing with the wide applications of nano-materials in mechanical, automobile, and aerospace industries. Various models, such as the free sliding model [1], the linear spring model [2], the dislocation-like model [3], the interphase model [4], the interface stress model [5-9] etc., are developed to simulate the mechanical properties of interfaces in nano-materials. Among these models, the Steigmann–Ogden interface stress model [8, 9] (hereinafter referred to as the S−O model) has enjoyed an increasing popularity. As both stretching resistance and the bending resistance are incorporated into the surface/interface constitutive relation, the S−O model can account for the known experimental observations and simulation results on the size-dependent surface stresses of nanowires [10, 11], nanoplates [12] and nanoparticles [13], which cannot be fully explained by the Gurtin–Murdoch model [6, 7].

The Steigmann–Ogden interface stress model was first put forward by Steigmann and Ogden[8] in 1997, and has recently attracted increased interest since Eremeyev and Lebedev[14, 15] derived equilibrium equations and boundary conditions describing an elastic solid with surface stresses. The interface can be regarded as a negligibly thin shell attached to the surface/interface of the bulk materials in the S–O model, while in the G–M model the surface/interface is regarded as a membrane capable only of stretching (no flexural resistance) leading to the possibility of instability under compressive surface/interface stresses (e.g. wrinkling) [16]. The S–O model is thus recognized as a advancement in the field of surface mechanics [14-17], and has been widely used for mechanical analysis of nano-materials, e.g. nanobeams [18], nanowires [19], nanoshells [20], polymer brush [21, 22] and half-space material [23, 24].

In contrast to the large number of studies available for materials with the Gurtin-Murdoch interface model (e.g. [25-39] and many others), the literature on nano-porous materials and nano-particle reinforced composites, considering the Steigmann–Ogden model, is rather limited. The only papers we are aware of are those by Dai et al. [40], Gharahi and Schiavone [16], Han et al. [41], Zemlyanova and Mogilevskaya [42]. Among these studies, Dai et al. [40] and Zemlyanova and Mogilevskaya [42] presented analytical/ semi-analytical solutions to the two-dimensional problem of an infinite isotropic elastic domain containing an isotropic elastic circular inhomogeneity; Gharahi and Schiavone [16] derived the effective moduli of the micropolar nano-composite considering plane elasticity.

In our previous study [17], we presented an explicit solution for the problem of a spherical nano-inhomogeneity embedded in an infinite matrix loaded by uniform far-field-stresses. It was shown that the existence of interface bending resistance can significantly change the local stress distributions around the interface. However, due to the mathematical complexity, studies of the effective properties of 3D nano-composites with the Steigmann–Ogden interface model have not been reported to the best of our knowledge. Especially, the explicit expressions of the Eshelby tensor and overall properties of nano-composites and nano-porous materials considering the Steigmann–Ogden interface model, which can be quite useful in the designing nanocomposites and porous materials, are very desirable.

Following our previous study [17], the Eshelby formalism is extended to the problem of a nano-inclusion with the Steigmann–Ogden interface model for the first time in this study. Using the obtained stress/strain concentration tensors, we employ the dilute approximation and the Mori-Tanakas method to derive the effective elastic moduli of nano-composites following the procedure given in Duan et al. [30]. The derived formulas can also be used for nano-porous materials by setting the moduli of the inclusion to be 0. An equivalent interface curvature parameter, and a characteristic curvature parameter are also defined, and their significances are discussed in detail.

The rest of this paper is organized as follows: In Section 2, the governing equations for the 3D nano-inhomogeneity with Steigmann–Ogden interface are briefly stated. In Section 3, explicit analytical solutions and Eshelby tensors for an inclusion subjected to eigenstrains are given. In Section 4, the explicit solutions and stress/strain concentration tensors for a spherical inhomogeneity with far-field stresses/strains are given. In Section 5, the expressions and numerical examples of the effective elastic moduli of nano-composites, considering the Steigmann–Ogden interface model, are given. In Section 6, we complete this paper with some concluding remarks.

**2. The governing linear elasticity equations**

We start by considering an infinite matrix with a nanosized inclusion/inhomogeneity illustrated in Fig. 1. Solutions for the matrix and the inhomogeneity should satisfy the equations of stress equilibrium, strain displacement-gradient compatibility, constitutive relations, as well as far-field boundary conditions.

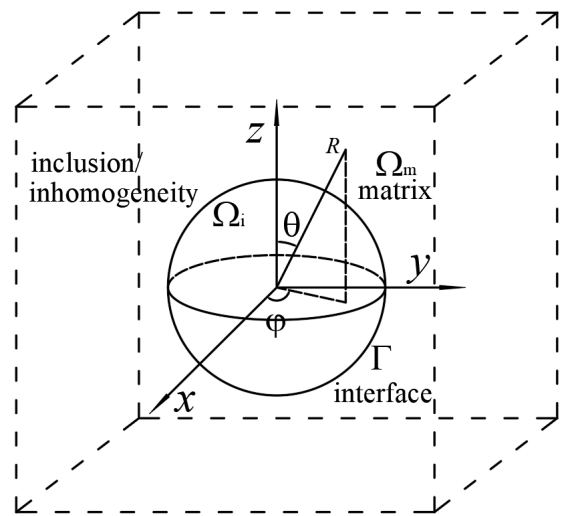

Fig. 1.    A nanosized spherical inhomogeneity embedded in an infinite matrix

The Steigmann–Ogden interface model is employed to consider the interface bending resistance as well as interface stretching resistance. Derivation of the Steigmann–Ogden interface model is detailed in [14, 15], and the governing equations are summarized below. The displacement vector across the interface is continuous,

$$\mathbf{u}_m = \mathbf{u}_i \quad \text{at } \Gamma \tag{1}$$

The stress tensor across the interface has a jump,

$$\mathbf{n} \cdot \Delta \boldsymbol{\sigma} = \nabla_s \cdot \left[ \boldsymbol{\tau}_s + (\nabla_s \cdot \mathbf{m}_s)\mathbf{n} \right] - (\nabla_s \cdot \mathbf{n})\mathbf{n} \cdot (\nabla_s \cdot \mathbf{m}_s)\mathbf{n} \quad \text{at } \Gamma \tag{2}$$

The constitutive equation of the interface is

$$\boldsymbol{\tau}_s = 2\mu_s \boldsymbol{\varepsilon}_s + \lambda_s \text{tr}(\boldsymbol{\varepsilon}_s) \mathbf{I}_s \tag{3}$$

$$\mathbf{m}_s = 2\chi_s \boldsymbol{\kappa}_s + \zeta_s \text{tr}(\boldsymbol{\kappa}_s) \mathbf{I}_s \tag{4}$$

where

$$\boldsymbol{\varepsilon}_s = \frac{1}{2}\left[ \nabla_s \mathbf{u}_s \cdot \mathbf{I}_s + \mathbf{I}_s \cdot (\nabla_s \mathbf{u}_s)^T \right] \tag{5}$$

$$\boldsymbol{\kappa}_s = -\frac{1}{2}\left[ \nabla_s \vartheta \cdot \mathbf{I}_s + \mathbf{I}_s \cdot (\nabla_s \vartheta)^T \right] \tag{6}$$

$$\vartheta = \nabla_s (\mathbf{n} \cdot \mathbf{u}_s) + \mathbf{B} \cdot \mathbf{u}_s \tag{7}$$

$$\mathbf{B} = -\nabla_s \mathbf{n} \tag{8}$$

in which $\mathbf{u}_s, \boldsymbol{\varepsilon}_s, \boldsymbol{\tau}_s, \boldsymbol{\kappa}_s$ and $\mathbf{m}_s$ are the interface fields of displacement, strain, stress, curvature and bending moment, respectively. $\lambda_s$ and $\mu_s$ are the stiffness parameters characterizing the interface stretching. $\chi_s$ and $\zeta_s$ are the stiffness parameters characterizing the interface bending. $\mathbf{I}_s$ is the unit tangent tensor defined on the interface and $\mathbf{I}_s = \mathbf{e}_\theta \otimes \mathbf{e}_\theta + \mathbf{e}_\varphi \otimes \mathbf{e}_\varphi$ in spherical coordinates. $\nabla_s = (\mathbf{I}_3 - \mathbf{nn}) \cdot \nabla$ is the gradient operator defined on the interface where $\mathbf{n}$ is the unit outer-normal vector of the interface $\Gamma$.

## 3. Eshelby tensor for a nano-inclusion

In this section, we consider a spherical inclusion of radius $R$ embedded in an infinite matrix, along with an interface characterized by the S–O model. The elastic modulus is the same for both the inclusion and the matrix ($E_m = E_i$). The eigenstrains $\boldsymbol{\varepsilon}^*$ in the inclusion are assumed to be uniform. For this problem, the analytical solution can be derived easily by employing Papkovich–Neuber solutions [43, 44]. We express the analytical solution for this problem as being a linear combination of Papkovich–Neuber potentials, and determine the unknown coefficients by enforcing the far-field and interface conditions. Details of the derivation can be found in Wang et al. [17].

### 3.1. Analytical solutions with uniform eigenstrains

First, we consider the case that the eigenstrain tensor $\boldsymbol{\varepsilon}^*$ has only one non-zero component $\varepsilon_{xx}^*$. The displacement components of $\mathbf{u}_{xx}^j$ (here we use $\mathbf{u}_{\alpha\beta}^j$ to denote the displacement vector when the eigenstrain has only one non-zero component $\varepsilon_{\alpha\beta}^*$) can be simplified as:

$$\begin{aligned}
u_{rxx}^m &= \frac{M_{1xx}}{r^2} + \frac{1}{8r^4}\left(3M_{3xx} + 2M_{2xx}r^2(-5+4v_m)\right)\left(1+3\cos[2\theta]-6\cos[2\varphi]\sin[\theta]^2\right); \\
u_{\theta xx}^m &= \frac{3\left(M_{3xx}+2M_{2xx}r^2(1-2v_m)\right)}{r^4}\cos[\theta]\cos[\varphi]^2\sin[\theta]; \\
u_{\varphi xx}^m &= -\frac{3\left(M_{3xx}+2M_{2xx}r^2(1-2v_m)\right)}{r^4}\cos[\varphi]\sin[\theta]\sin[\varphi]; \\
u_{rxx}^i &= C_{1xx}r - \frac{1}{4}r\left(C_{3xx}+6C_{2xx}r^2 v_m\right)\left(1+3\cos[2\theta]-6\cos[2\varphi]\sin[\theta]^2\right); \\
u_{\theta xx}^i &= 3\left(C_{3xx}r + C_{2xx}r^3(7-4v_m)\right)\cos[\theta]\cos[\varphi]^2\sin[\theta]; \\
u_{\varphi xx}^i &= -3\left(C_{3xx}r + C_{2xx}r^3(7-4v_m)\right)\cos[\varphi]\sin[\theta]\sin[\varphi];
\end{aligned} \quad (9)$$

By using the same procedure, the displacement field $\mathbf{u}_{yy}^j$ with eigenstrain $\varepsilon_{yy}^*$ can

be written as:

$$u_{ryy}^m = \frac{M_{1yy}}{r^2} + \frac{1}{8r^4}\left(3M_{3yy} + 2M_{2yy}r^2(-5+4v_m)\right)\left(1+3\cos[2\theta]+6\cos[2\varphi]\sin[\theta]^2\right);$$

$$u_{\theta yy}^m = \frac{3\left(M_{3yy} + 2M_{2yy}r^2(1-2v_m)\right)\cos[\theta]\sin[\theta]\sin[\varphi]^2}{r^4};$$

$$u_{\varphi yy}^m = \frac{3\left(M_{3yy} + 2M_{2yy}r^2(1-2v_m)\right)\cos[\varphi]\sin[\theta]\sin[\varphi]}{r^4}; \quad (10)$$

$$u_{ryy}^i = C_{1yy}r - \frac{1}{4}r\left(C_{3yy} + 6C_{2yy}r^2 v_m\right)\left(1+3\cos[2\theta]+6\cos[2\varphi]\sin[\theta]^2\right);$$

$$u_{\theta yy}^i = 3\left(C_{3yy}r + C_{2yy}r^3(7-4v_m)\right)\cos[\theta]\sin[\theta]\sin[\varphi]^2;$$

$$u_{\varphi yy}^i = 3\left(C_{3yy}r + C_{2yy}r^3(7-4v_m)\right)\cos[\varphi]\sin[\theta]\sin[\varphi];$$

The displacement components of $\mathbf{u}_{zz}^j$ with eigenstrain $\varepsilon_{zz}^*$ can be written as:

$$u_{rzz}^m = \frac{M_{1zz}}{r^2} - \frac{\left(3M_{3zz} + 2M_{2zz}r^2(-5+4v_m)\right)\left(1+3\cos[2\theta]\right)}{4r^4};$$

$$u_{\theta zz}^m = -\frac{3\left(M_{3zz} + 2M_{2zz}r^2(1-2v_m)\right)\cos[\theta]\sin[\theta]}{r^4};$$

$$u_{\varphi zz}^m = 0; \quad (11)$$

$$u_{rzz}^i = C_{1zz}r + \frac{1}{2}r\left(C_{3zz} + 6C_{2zz}r^2 v_m\right)\left(1+3\cos[2\theta]\right);$$

$$u_{\theta zz}^i = -3\left(C_{3zz}r + C_{2zz}r^3(7-4v_m)\right)\cos[\theta]\sin[\theta];$$

$$u_{\varphi zz}^i = 0;$$

The displacement field $\mathbf{u}_{xy}^j$ with eigenstrain $\varepsilon_{xy}^*$ can be written as:

$$u_{rxy}^m = \frac{M_{1xy}}{r^2} - \frac{3\left(3M_{3xy} + 2M_{2xy}r^2(-5+4v_m)\right)\sin[\theta]^2\sin[2\varphi]}{2r^4};$$

$$u_{\theta xy}^m = \frac{3\left(M_{3xy} + 2M_{2xy}r^2(1-2v_m)\right)\sin[2\theta]\sin[2\varphi]}{2r^4};$$

$$u_{\varphi xy}^m = \frac{3\left(M_{3xy} + 2M_{2xy}r^2(1-2v_m)\right)\cos[2\varphi]\sin[\theta]}{r^4}; \quad (12)$$

$$u_{rxy}^i = r\left(C_{1xy} + 3\left(C_{3xy} + 6C_{2xy}r^2 v_m\right)\sin[\theta]^2\sin[2\varphi]\right);$$

$$u_{\theta xy}^i = \frac{3}{2}r\left(C_{3xy} + C_{2xy}r^2(7-4v_m)\right)\sin[2\theta]\sin[2\varphi];$$

$$u_{\varphi xy}^i = 3\left(C_{3xy}r + C_{2xy}r^3(7-4v_m)\right)\cos[2\varphi]\sin[\theta];$$

The displacement field $\mathbf{u}_{yz}^j$ with eigenstrain $\varepsilon_{yz}^*$ can be written as:

$$u_{ryz}^{m} = \frac{M_{1yz}}{r^{2}} - \frac{3\left(3M_{3yz} + 2M_{2yz}r^{2}(-5+4v_{m})\right)\sin[2\theta]\sin[\varphi]}{2r^{4}};$$

$$u_{\theta yz}^{m} = \frac{3\left(M_{3yz} + 2M_{2yz}r^{2}(1-2v_{m})\right)\cos[2\theta]\sin[\varphi]}{r^{4}};$$

$$u_{\varphi yz}^{m} = \frac{3\left(M_{3yz} + 2M_{2yz}r^{2}(1-2v_{m})\right)\cos[\theta]\cos[\varphi]}{r^{4}}; \quad (13)$$

$$u_{ryz}^{i} = r\left(C_{1yz} + 3\left(C_{3yz} + 6C_{2yz}r^{2}v_{m}\right)\sin[2\theta]\sin[\varphi]\right);$$

$$u_{\theta yz}^{i} = 3\left(C_{3yz}r + C_{2yz}r^{3}(7-4v_{m})\right)\cos[2\theta]\sin[\varphi];$$

$$u_{\varphi yz}^{i} = 3\left(C_{3yz}r + C_{2yz}r^{3}(7-4v_{m})\right)\cos[\theta]\cos[\varphi];$$

The displacement field $\mathbf{u}_{zx}^{j}$ with eigenstrain $\varepsilon_{zx}^{*}$ can be written as:

$$u_{rzx}^{m} = \frac{M_{1zx}}{r^{2}} - \frac{3\left(3M_{3zx} + 2M_{2zx}r^{2}(-5+4v_{m})\right)\cos[\varphi]\sin[2\theta]}{2r^{4}};$$

$$u_{\theta zx}^{m} = \frac{3\left(M_{3zx} + 2M_{2zx}r^{2}(1-2v_{m})\right)\cos[2\theta]\cos[\varphi]}{r^{4}};$$

$$u_{\varphi zx}^{m} = -\frac{3\left(M_{3zx} + 2M_{2zx}r^{2}(1-2v_{m})\right)\cos[\theta]\sin[\varphi]}{r^{4}}; \quad (14)$$

$$u_{rzx}^{i} = r\left(C_{1zx} + 3\left(C_{3zx} + 6C_{2zx}r^{2}v_{m}\right)\cos[\varphi]\sin[2\theta]\right);$$

$$u_{\theta zx}^{i} = 3\left(C_{3zx}r + C_{2zx}r^{3}(7-4v_{m})\right)\cos[2\theta]\cos[\varphi];$$

$$u_{\varphi zx}^{i} = -3\left(C_{3zx}r + C_{2zx}r^{3}(7-4v_{m})\right)\cos[\theta]\sin[\varphi];$$

where $M_{p\alpha\beta}(p=1,...,5 \text{ and } \alpha,\beta=x,y,z)$ and $C_{q\alpha\beta}(q=1,2,3 \text{ and } \alpha,\beta=x,y,z)$ are constants given in the Appendix.

On the above, we have obtained the basic solutions for a spherical inclusion with six different eigenstrains. Thus, the analytical solution with an arbitrary uniform eigenstrain $\varepsilon^{*}$ can be expressed as an additive combination of Eqs.(9-14):

$$\mathbf{u}^{j} = \mathbf{u}_{xx}^{j} + \mathbf{u}_{yy}^{j} + \mathbf{u}_{zz}^{j} + \mathbf{u}_{xy}^{j} + \mathbf{u}_{yz}^{j} + \mathbf{u}_{zx}^{j} \quad j = m,i \quad (15)$$

Using strain displacement-gradient compatibility and the constitutive relations, the strain/stress fields can be obtained.

## 3.2. Eshelby tensors for the matrix and the inclusion

Using the obtained solutions in the previous subsection, we can calculate the Eshelby tensor. Unlike the classical interior Eshelby tensor, the Eshelby tensor in the case of the nano-inclusion is no longer uniform due to interface stress effects. Using the Walpole notation [45], a fourth-order tensor $\mathbf{S}^k(r)$ with radial symmetry can be expressed as

$$\mathbf{S}^k(r) = S_1^k(r)\mathbf{E}^1 + S_2^k(r)\mathbf{E}^2 + S_3^k(r)\mathbf{E}^3 + S_4^k(r)\mathbf{E}^4 + S_5^k(r)\mathbf{E}^5 + S_6^k(r)\mathbf{E}^6 \tag{16}$$

where $\mathbf{E}^i$ are elementary tensors introduced by Walpole [45]:

$$\begin{aligned}
E_{ijkl}^1 &= \frac{1}{2} b_{ij} b_{kl} \\
E_{ijkl}^2 &= a_{ij} a_{kl} \\
E_{ijkl}^3 &= \frac{1}{2}\left(b_{ik} b_{jl} + b_{jk} b_{il} - b_{ij} b_{kl}\right) \\
E_{ijkl}^4 &= \frac{1}{2}\left(b_{ik} a_{jl} + b_{il} a_{jk} + b_{jl} a_{ik} + b_{jk} a_{il}\right) \\
E_{ijkl}^5 &= a_{ij} b_{kl} \\
E_{ijkl}^6 &= b_{ij} a_{kl}
\end{aligned} \tag{17}$$

where $a_{ij}$ and $b_{ij}$ are components of $\mathbf{a} = \mathbf{e}_r \otimes \mathbf{e}_r$ and $\mathbf{b} = \mathbf{I}_2 - \mathbf{e}_r \otimes \mathbf{e}_r$, respectively.

In the inclusion, the Eshelby tensor can be written as:

$$\begin{aligned}
S_1^i &= 2C_1 + C_3 + 3C_2 r^2 (7 - 8v_m) \\
S_2^i &= C_1 + 2C_3 + 36 C_2 r^2 v_m \\
S_3^i &= 3\left(C_3 + C_2 r^2 (7 - 4v_m)\right) \\
S_4^i &= 3\left(C_3 + C_2 r^2 (7 + 2v_m)\right) \\
S_5^i &= C_1 - C_3 - 18 C_2 r^2 v_m \\
S_6^i &= C_1 - C_3 + 3 C_2 r^2 (-7 + 8v_m)
\end{aligned} \tag{18}$$

In the matrix, the Eshelby tensor can be written as:

$$S_1^m = \frac{2(3M_3 + r^2(M_1 - 2M_2(1+v_m)))}{r^5}$$

$$S_2^m = -\frac{2(-6M_3 + r^2(M_1 + 2M_2(5-4v_m)))}{r^5}$$

$$S_3^m = \frac{3(M_3 + 2M_2 r^2(1-2v_m))}{r^5}$$ (19)

$$S_4^m = \frac{6(-2M_3 + M_2 r^2(1+v_m))}{r^5}$$

$$S_5^m = -\frac{2(3M_3 + r^2(M_1 + M_2(-5+4v_m)))}{r^5}$$

$$S_6^m = \frac{-6M_3 + r^2(M_1 + 4M_2(1+v_m))}{r^5}$$

where $C_1, C_2, C_3, M_1, M_2, M_3$ are constants given in the Appendix.

The average Eshelby tensor in the inclusion is thus defined as:

$$\overline{\mathbf{S}}^i = \frac{1}{V_i} \int_{V_i} \mathbf{S}^i \mathrm{d}V$$ (20)

in which $V_i$ is the volume of the inclusion. After some derivations, it can be shown that $\overline{\mathbf{S}}^i$ is an isotropic tensor:

$$\overline{\mathbf{S}}^i = 3C_1 \mathbf{J} + \left(3C_3 + \frac{63C_2 R^2}{5}\right)\mathbf{K}$$

$$\mathbf{J} = \frac{1}{3}\mathbf{I}_2 \otimes \mathbf{I}_2, \mathbf{K} = \mathbf{I}_{4s} - \mathbf{J}$$ (21)

The solution (Eqs.(16-21)) reveals that the Eshelby tensor is not uniform and is affected by the interface properties. If the interface stress is neglected, Eqs.(16-21) degenerate into the classical Eshelby tensor without interface stress effects, and the size effect will no longer exist. If the surface bending resistance is neglected ($\chi_s = 0$ and $\zeta_s = 0$), Eqs.(16-21) degenerate into the Eshelby tensor considering the Gurtin-Murdoch interface model as given in Duan et al. [30].

## 4. Stress/strain concentration tensors for a nano-inhomogeneity

In this section, we consider a spherical inhomogeneity of radius $R$ embedded in an infinite matrix, with the S-O interface, subjected to homogeneous far-field stresses and strains at infinity. The far-field condition for uniform stresses and strains can be written as:

$$\boldsymbol{\sigma}_m = \boldsymbol{\Sigma} \quad \text{at infinity} \tag{22}$$

$$\boldsymbol{\varepsilon}_m = \mathbf{E} \quad \text{at infinity} \tag{23}$$

where $\boldsymbol{\Sigma}$ and $\mathbf{E}$ are constant far-field strain and stress tensors. It should be pointed out that the equivalent inclusion method is not applicable to this problem, because the strain field in the inhomogeneity is not uniform. We will discuss these two far-field conditions in the following two subsections separately. The detailed derivation of the displacement fields for this problem is similar to Wang et al. [17], thus we simply list the results in the present paper.

*4.1. Stress concentration tensors*

First, we consider the case when the remote loading has only one non-zero stress component $\Sigma_{xx}$. The displacement field $\mathbf{u}^j_{xx}$ (here we use $\mathbf{u}^j_{\alpha\beta}$ to denote the displacement vector when remote loading has only one non-zero component $\Sigma_{\alpha\beta}$) can be simplified as:

$$u_{rxx}^m = \frac{M_{2xx}^\Sigma}{r^2} + M_{1xx}^\Sigma r - \frac{1}{8}\left(-\frac{3M_{5xx}^\Sigma}{r^4} + 2M_{3xx}^\Sigma r + \frac{2M_{4xx}^\Sigma(5-4v_m)}{r^2}\right)$$

$$\left(1 + 3\cos[2\theta] - 6\cos[2\varphi]\sin[\theta]^2\right);$$

$$u_{\theta xx}^m = \frac{3}{r^4}\left(M_{5xx}^\Sigma + M_{3xx}^\Sigma r^5 + M_{4xx}^\Sigma r^2(2-4v_m)\right)\cos[\theta]\cos[\varphi]^2\sin[\theta];$$

$$u_{\varphi xx}^m = -\frac{3}{r^4}\left(M_{5xx}^\Sigma + M_{3xx}^\Sigma r^5 + M_{4xx}^\Sigma r^2(2-4v_m)\right)\cos[\varphi]\sin[\theta]\sin[\varphi]; \quad (24)$$

$$u_{rxx}^i = C_{1xx}^\Sigma r - \frac{1}{4}r\left(C_{3xx}^\Sigma + 6C_{2xx}^\Sigma r^2 v_i\right)\left(1 + 3\cos[2\theta] - 6\cos[2\varphi]\sin[\theta]^2\right);$$

$$u_{\theta xx}^i = 3\left(C_{3xx}^\Sigma r + C_{2xx}^\Sigma r^3(7-4v_i)\right)\cos[\theta]\cos[\varphi]^2\sin[\theta];$$

$$u_{\varphi xx}^i = -3\left(C_{3xx}^\Sigma r + C_{2xx}^\Sigma r^3(7-4v_i)\right)\cos[\varphi]\sin[\theta]\sin[\varphi];$$

The displacement field $\mathbf{u}_{yy}^j$ with the remote tensile stress $\Sigma_{yy}$ can be written as:

$$u_{ryy}^m = \frac{M_{2yy}^\Sigma}{r^2} + M_{1yy}^\Sigma r - \frac{1}{8}\left(-\frac{3M_{5yy}^\Sigma}{r^4} + 2M_{3yy}^\Sigma r + \frac{2M_{4yy}^\Sigma(5-4v_m)}{r^2}\right)$$

$$\left(1 + 3\cos[2\theta] + 6\cos[2\varphi]\sin[\theta]^2\right);$$

$$u_{\theta yy}^m = \frac{3}{r^4}\left(M_{5yy}^\Sigma + M_{3yy}^\Sigma r^5 + M_{4yy}^\Sigma r^2(2-4v_m)\right)\cos[\theta]\sin[\theta]\sin[\varphi]^2;$$

$$u_{\varphi yy}^m = \frac{3}{r^4}\left(M_{5yy}^\Sigma + M_{3yy}^\Sigma r^5 + M_{4yy}^\Sigma r^2(2-4v_m)\right)\cos[\varphi]\sin[\theta]\sin[\varphi]; \quad (25)$$

$$u_{ryy}^i = C_{1yy}^\Sigma r - \frac{1}{4}r\left(C_{3yy}^\Sigma + 6C_{2yy}^\Sigma r^2 v_i\right)\left(1 + 3\cos[2\theta] + 6\cos[2\varphi]\sin[\theta]^2\right);$$

$$u_{\theta yy}^i = 3\left(C_{3yy}^\Sigma r + C_{2yy}^\Sigma r^3(7-4v_i)\right)\cos[\theta]\sin[\theta]\sin[\varphi]^2;$$

$$u_{\varphi yy}^i = 3\left(C_{3yy}^\Sigma r + C_{2yy}^\Sigma r^3(7-4v_i)\right)\cos[\varphi]\sin[\theta]\sin[\varphi];$$

The displacement field $\mathbf{u}_{zz}^j$ with the remote tensile stress $\Sigma_{zz}$ can be written as:

$$u_{rzz}^m = \frac{M_{2zz}^\Sigma}{r^2} + M_{1zz}^\Sigma r + \frac{1}{4}\left(-\frac{3M_{5zz}^\Sigma}{r^4} + 2M_{3zz}^\Sigma r + \frac{2M_{4zz}^\Sigma(5-4v_m)}{r^2}\right)\left(1 + 3\cos[2\theta]\right);$$

$$u_{\theta zz}^m = -\frac{3}{r^4}\left(M_{5zz}^\Sigma + M_{3zz}^\Sigma r^5 + M_{4zz}^\Sigma r^2(2-4v_m)\right)\cos[\theta]\sin[\theta];$$

$$u_{\varphi zz}^m = 0; \quad (26)$$

$$u_{rzz}^i = C_{1zz}^\Sigma r + \frac{1}{2}r\left(C_{3zz}^\Sigma + 6C_{2zz}^\Sigma r^2 v_i\right)\left(1 + 3\cos[2\theta]\right);$$

$$u_{\theta zz}^i = -3\left(C_{3zz}^\Sigma r + C_{2zz}^\Sigma r^3(7-4v_i)\right)\cos[\theta]\sin[\theta];$$

$$u_{\varphi zz}^i = 0;$$

The displacement field $\mathbf{u}_{xy}^j$ with the remote shear stress $\Sigma_{xy}$ can be written as:

$$u_{rxy}^m = \frac{M_{2xy}^\Sigma}{r^2} + \frac{3}{2}\left(-\frac{3M_{5xy}^\Sigma}{r^4} + 2M_{3xy}^\Sigma r + \frac{2M_{4xy}^\Sigma(5-4v_m)}{r^2}\right)\sin[\theta]^2\sin[2\varphi];$$

$$u_{\theta xy}^m = \frac{3}{2r^4}\left(M_{5xy}^\Sigma + M_{3xy}^\Sigma r^5 + M_{4xy}^\Sigma r^2(2-4v_m)\right)\sin[2\theta]\sin[2\varphi];$$

$$u_{\varphi xy}^m = \frac{3}{r^4}\left(M_{5xy}^\Sigma + M_{3xy}^\Sigma r^5 + M_{4xy}^\Sigma r^2(2-4v_m)\right)\cos[2\varphi]\sin[\theta]; \qquad (27)$$

$$u_{rxy}^i = r\left(C_{1xy}^\Sigma + 3\left(C_{3xy}^\Sigma + 6C_{2xy}^\Sigma r^2 v_i\right)\sin[\theta]^2\sin[2\varphi]\right);$$

$$u_{\theta xy}^i = \frac{3}{2}r\left(C_{3xy}^\Sigma + C_{2xy}^\Sigma r^2(7-4v_i)\right)\sin[2\theta]\sin[2\varphi];$$

$$u_{\varphi xy}^i = 3\left(C_{3xy}^\Sigma r + C_{2xy}^\Sigma r^3(7-4v_i)\right)\cos[2\varphi]\sin[\theta];$$

The displacement field $\mathbf{u}_{yz}^j$ with the remote shear stress $\Sigma_{yz}$ can be written as:

$$u_{ryz}^m = \frac{M_{2yz}^\Sigma}{r^2} + \frac{3}{2}\left(-\frac{3M_{5yz}^\Sigma}{r^4} + 2M_{3yz}^\Sigma r + \frac{2M_{4yz}^\Sigma(5-4v_m)}{r^2}\right)\sin[2\theta]\sin[\varphi];$$

$$u_{\theta yz}^m = \frac{3}{r^4}\left(M_{5yz}^\Sigma + M_{3yz}^\Sigma r^5 + M_{4yz}^\Sigma r^2(2-4v_m)\right)\cos[2\theta]\sin[\varphi];$$

$$u_{\varphi yz}^m = \frac{3}{r^4}\left(M_{5yz}^\Sigma + M_{3yz}^\Sigma r^5 + M_{4yz}^\Sigma r^2(2-4v_m)\right)\cos[\theta]\cos[\varphi]; \qquad (28)$$

$$u_{ryz}^i = r\left(C_{1yz}^\Sigma + 3\left(C_{3yz}^\Sigma + 6C_{2yz}^\Sigma r^2 v_i\right)\sin[2\theta]\sin[\varphi]\right);$$

$$u_{\theta yz}^i = 3\left(C_{3yz}^\Sigma r + C_{2yz}^\Sigma r^3(7-4v_i)\right)\cos[2\theta]\sin[\varphi];$$

$$u_{\varphi yz}^i = 3\left(C_{3yz}^\Sigma r + C_{2yz}^\Sigma r^3(7-4v_i)\right)\cos[\theta]\cos[\varphi];$$

The displacement field $\mathbf{u}_{zx}^j$ with the remote shear stress $\Sigma_{xy}$ can be written as:

$$u_{rzx}^m = \frac{M_{2zx}^\Sigma}{r^2} + \frac{3}{2}\left(-\frac{3M_{5zx}^\Sigma}{r^4} + 2M_{3zx}^\Sigma r + \frac{2M_{4zx}^\Sigma(5-4v_m)}{r^2}\right)\cos[\varphi]\sin[2\theta];$$

$$u_{\theta zx}^m = \frac{3}{r^4}\left(M_{5zx}^\Sigma + M_{3zx}^\Sigma r^5 + M_{4zx}^\Sigma r^2(2-4v_m)\right)\cos[2\theta]\cos[\varphi];$$

$$u_{\varphi zx}^m = -\frac{3}{r^4}\left(M_{5zx}^\Sigma + M_{3zx}^\Sigma r^5 + M_{4zx}^\Sigma r^2(2-4v_m)\right)\cos[\theta]\sin[\varphi]; \qquad (29)$$

$$u_{rzx}^i = r\left(C_{1zx}^\Sigma + 3\left(C_{3zx}^\Sigma + 6C_{2zx}^\Sigma r^2 v_i\right)\cos[\varphi]\sin[2\theta]\right);$$

$$u_{\theta zx}^i = 3\left(C_{3zx}^\Sigma r + C_{2zx}^\Sigma r^3(7-4v_i)\right)\cos[2\theta]\cos[\varphi];$$

$$u_{\varphi zx}^i = -3\left(C_{3zx}^\Sigma r + C_{2zx}^\Sigma r^3(7-4v_i)\right)\cos[\theta]\sin[\varphi];$$

where $M_{p\alpha\beta}^\Sigma (p=1,...,5 \text{ and } \alpha,\beta=x,y,z)$ and $C_{q\alpha\beta}^\Sigma (q=1,2,3 \text{ and } \alpha,\beta=x,y,z)$ are constants given in the Appendix. Now we have obtained the basic solutions for a spherical inhomogeneity under six different remote loading cases. Thus, the analytical solution under remote loading $\Sigma$ can be written as:

$$\mathbf{u}^j = \mathbf{u}^j_{xx} + \mathbf{u}^j_{yy} + \mathbf{u}^j_{zz} + \mathbf{u}^j_{xy} + \mathbf{u}^j_{yz} + \mathbf{u}^j_{zx} \quad j=m,i \tag{30}$$

Using strain displacement compatibility and the constitutive relations, the strain/stress fields can be obtained. Once the stress fields in the inhomogeneity and the matrix are derived, we can derive the stress concentration tensor for the inhomogeneity and the matrix considering the S-O model. The stress concentration tensors for the spherical inhomogeneity can be expressed as:

$$\mathbf{B}^k(r) = B_1^k(r)\mathbf{E}^1 + B_2^k(r)\mathbf{E}^2 + B_3^k(r)\mathbf{E}^3 + B_4^k(r)\mathbf{E}^4 + B_5^k(r)\mathbf{E}^5 + B_6^k(r)\mathbf{E}^6 \tag{31}$$

In the inhomogeneity, the stress concentration tensor can be written as:

$$\begin{aligned}
B_1^i &= 6C_2^\Sigma r^2 (7+6v_i)\mu_i + 2\left(C_3^\Sigma - \frac{2C_1^\Sigma(1+v_i)}{-1+2v_i}\right)\mu_i; \\
B_2^i &= 4C_3^\Sigma \mu_i - 12C_2^\Sigma r^2 v_i \mu_i - \frac{2C_1^\Sigma(1+v_i)\mu_i}{-1+2v_i}; \\
B_3^i &= 6C_3^\Sigma \mu_i + 6C_2^\Sigma r^2 (7-4v_i)\mu_i; \\
B_4^i &= 6C_3^\Sigma \mu_i + 6C_2^\Sigma r^2 (7+2v_i)\mu_i; \\
B_5^i &= 6C_2^\Sigma r^2 v_i \mu_i + 2\left(-C_3^\Sigma + \frac{C_1^\Sigma(1+v_i)}{1-2v_i}\right)\mu_i; \\
B_6^i &= -2C_3^\Sigma \mu_i - \frac{2C_1^\Sigma(1+v_i)\mu_i}{-1+2v_i} - 6C_2^\Sigma r^2 (7+6v_i)\mu_i;
\end{aligned} \tag{32}$$

In the matrix, the stress concentration tensor can be written as:

$$B_1^m = \frac{12M_5^\Sigma \mu_m}{r^5} + \frac{4(M_2^\Sigma - 2M_4^\Sigma + 4M_4^\Sigma v_m)\mu_m}{r^3} - \frac{2(M_3^\Sigma - 2M_3^\Sigma v_m + 2M_1^\Sigma(1+v_m))\mu_m}{-1+2v_m}$$

$$B_2^m = \frac{24M_5\mu_m}{r^5} - \frac{4(M_2 - 2M_4(-5+v_m))\mu_m}{r^3} - \frac{2(2M_3(1-2v_m) + M_1(1+v_m))\mu_m}{-1+2v_m}$$

$$B_3^m = 6M_3^\Sigma \mu_m + \frac{6M_5^\Sigma \mu_m}{r^5} + \frac{6M_4^\Sigma(2-4v_m)\mu_m}{r^3}$$

$$B_4^m = 6M_3^\Sigma \mu_m - \frac{24M_5^\Sigma \mu_m}{r^5} + \frac{12M_4^\Sigma(1+v_m)\mu_m}{r^3}$$

$$B_5^m = -\frac{12M_5^\Sigma \mu_m}{r^5} - \frac{4(M_2^\Sigma + M_4^\Sigma(-5+v_m))\mu_m}{r^3} - \frac{2(M_1^\Sigma - M_3^\Sigma + M_1^\Sigma v_m + 2M_3^\Sigma v_m)\mu_m}{-1+2v_m}$$

$$B_6^m = -\frac{12M_5^\Sigma \mu_m}{r^5} + \frac{2(M_2^\Sigma + M_4^\Sigma(4-8v_m))\mu_m}{r^3} - \frac{2(M_1^\Sigma - M_3^\Sigma + M_1^\Sigma v_m + 2M_3^\Sigma v_m)\mu_m}{-1+2v_m}$$

(33)

where $C_1^\Sigma, C_2^\Sigma, C_3^\Sigma, M_1^\Sigma, M_2^\Sigma, M_3^\Sigma, M_4^\Sigma, M_5^\Sigma$ are constants given in the Appendix.

The average stress concentration tensor in the inhomogeneity is defined as:

$$\bar{\mathbf{B}}^i = \frac{1}{V_i}\int_{V_i} \mathbf{B}^i dV \quad (34)$$

Again after some manipulations, it can be shown that $\bar{\mathbf{B}}$ is an isotropic tensor:

$$\bar{\mathbf{B}}^i = \bar{B}_1^i \mathbf{J} + \bar{B}_2^i \mathbf{K}$$
$$\bar{B}_1^i = -\frac{6C_1^\Sigma(1+v_i)\mu_i}{-1+2v_i} \quad (35)$$
$$\bar{B}_2^i = \frac{6}{5}(5C_3^\Sigma + 21C_2^\Sigma R^2)\mu_i$$

where $\mathbf{J} = \frac{1}{3}\mathbf{I}_2 \otimes \mathbf{I}_2$ and $\mathbf{K} = \mathbf{I}_{4s} - \mathbf{J}$. $\mathbf{I}_2$ is the second-order identity tensor and $\mathbf{I}_{4s}$ is the fourth-order symmetric identity tensor. As seen from Eq.(35), the average stress concentration tensor in the inhomogeneity is an isotropic tensor of fourth order.

In order to derive the effective compliance tensor in the next section, we introduce the interface stress concentration tensors, which is defined as below:

$$\frac{1}{V_i}\int_\Gamma (\boldsymbol{\sigma}^m - \boldsymbol{\sigma}^i)\cdot \mathbf{n} \otimes \mathbf{r} dS = \bar{\mathbf{B}}^\Gamma : \Sigma \quad (36)$$

where

$$\bar{\mathbf{B}}^{\Gamma} = \bar{B}_1^{\Gamma}\mathbf{J} + \bar{B}_2^{\Gamma}\mathbf{K}$$

$$\bar{B}_1^{\Gamma} = \frac{6C_1^{\Sigma}(1+v_i)\mu_i}{-1+2v_i} - \frac{6(M_1^{\Sigma}R^3(1+v_m) + M_2^{\Sigma}(-2+4v_m))\mu_m}{R^3(-1+2v_m)} \quad (37)$$

$$\bar{B}_2^{\Gamma} = \frac{6}{5}\left(-5C_3^{\Sigma}\mu_i - 21C_2^{\Sigma}R^2\mu_i + 5M_3^{\Sigma}\mu_m + \frac{2M_4^{\Sigma}(-7+5v_m)\mu_m}{R^3}\right)$$

In addition to size-dependency, two interesting phenomena are observed from the closed form Eqs.(24-37). First, displacements and stress concentration tensors depend on the interface curvature parameter:

$$\eta_s = 3\zeta_s + 5\chi_s \quad (38)$$

that is to say, they are not affected individually by $\zeta_s$ and $\chi_s$ when $\eta_s$ is a fixed value. Second, the stress concentration tensor in the inhomogeneity is transversely isotropic, while the average stress concentration tensor is an isotropic tensor.

### 4.2. Strain concentration tensors

The analytical solutions with far-field strain loading have the same structure as the analytical solutions with far-field stresses, except that the coefficients are different. The coefficients $M^E_{p\alpha\beta}(p=1,\ldots,5 \text{ and } \alpha,\beta = x,y,z)$ and $C^E_{q\alpha\beta}(q=1,2,3 \text{ and } \alpha,\beta = x,y,z)$ for uniform strain far-field condition can be found in the Appendix.

The strain concentration tensors for the spherical inhomogeneity can be expressed as:

$$\mathbf{A}^k(r) = A_1^k(r)\mathbf{E}^1 + A_2^k(r)\mathbf{E}^2 + A_3^k(r)\mathbf{E}^3 + A_4^k(r)\mathbf{E}^4 + A_5^k(r)\mathbf{E}^5 + A_6^k(r)\mathbf{E}^6 \quad (39)$$

In the inhomogeneity, the strain concentration tensor can be written as:

17 of 34 pages

$$A_1^i = 2C_1^E + C_3^E + 3C_2^E r^2 (7 - 8v_i);$$
$$A_2^i = C_1^E + 2C_3^E + 36C_2^E r^2 v_i;$$
$$A_3^i = 3C_3^E + 3C_2^E r^2 (7 - 4v_i);$$
$$A_4^i = 3C_3^E + 3C_2^E r^2 (7 + 2v_i); \quad (40)$$
$$A_5^i = C_1^E - C_3^E - 18C_2^E r^2 v_i;$$
$$A_6^i = C_1^E - C_3^E + 3C_2^E r^2 (-7 + 8v_i);$$

In the matrix, the strain concentration tensor can be written as:

$$A_1^m = 2M_1^E + M_3^E + \frac{6M_5^E}{r^5} + \frac{2M_2^E - 4M_4^E(1+v_m)}{r^3}$$

$$A_2^m = M_1^E + 2M_3^E + \frac{12M_5^E}{r^5} - \frac{2(M_2^E + 2M_4^E(5-4v_m))}{r^3}$$

$$A_3^m = 3M_3^E + \frac{3M_5^E}{r^5} + \frac{3M_4^E(2-4v_m)}{r^3}$$

$$A_4^m = 3M_3^E - \frac{12M_5^E}{r^5} + \frac{6M_4^E(1+v_m)}{r^3} \quad (41)$$

$$A_5^m = M_1^E - M_3^E - \frac{6M_5^E}{r^5} - \frac{2(M_2^E + M_4^E(-5+4v_m))}{r^3}$$

$$A_6^m = M_1^E - M_3^E - \frac{6M_5^E}{r^5} + \frac{M_2^E + 4M_4^E(1+v_m)}{r^3}$$

where $C_1^E, C_2^E, C_3^E, M_1^E, M_2^E, M_3^E, M_4^E, M_5^E$ are constants given in the Appendix.

The average strain concentration tensor in the inhomogeneity is defined as:

$$\overline{\mathbf{A}}^i = \frac{1}{V_i} \int_{V_i} \mathbf{A}^i dV \quad (42)$$

in which $V_i$ is the volume of the inhomogeneity. Again, $\overline{\mathbf{A}}^i$ is an isotropic tensor:

$$\overline{\mathbf{A}}^i = \overline{A}_1^i \mathbf{J} + \overline{A}_2^i \mathbf{K}$$
$$\overline{A}_1^i = 3C_1^E \quad (43)$$
$$\overline{A}_2^i = 3C_3^E + \frac{63C_2^E R^2}{5}$$

We also introduce the "interface strain concentration tensors"[31], which is defined as follows:

$$\frac{1}{V_i} \int_\Gamma (\boldsymbol{\sigma}^m - \boldsymbol{\sigma}^i) \cdot \mathbf{n} \otimes \mathbf{r} = \overline{\mathbf{A}}^\Gamma : \mathbf{E} \quad (44)$$

where

$$\overline{\mathbf{A}}^{\Gamma} = \overline{A}_1^{\Gamma}\mathbf{J} + \overline{A}_2^{\Gamma}\mathbf{K}$$
$$\overline{A}_1^{\Gamma} = \frac{6C_1^E(1+v_i)\mu_i}{-1+2v_i} - \frac{12M_2^E\mu_m}{R^3} - \frac{6M_1^E(1+v_m)\mu_m}{-1+2v_m} \quad (45)$$
$$\overline{A}_2^{\Gamma} = -6C_3^E\mu_i - \frac{126}{5}C_2^E R^2\mu_i + 6M_3^E\mu_m + \frac{12M_4^E(-7+5v_m)\mu_m}{5R^3}$$

Similar to the solution for far-field stresses, the solutions in Eqs.(39-45) depend on the interface curvature parameter $\eta_s = 3\zeta_s + 5\chi_s$. Besides, the strain concentration tensor in the inhomogeneity is transversely isotropic, while the average strain concentration tensor is an isotropic tensor.

## 5. Effective bulk and shear moduli

### 5.1. The dilute approximation and the Mori-Tanaka method

The effective stiffness tensor $\mathbf{L}^{hom}$ and compliance tensor $\mathbf{M}^{hom}$ are respectively defined as

$$\overline{\boldsymbol{\sigma}} = \mathbf{L}^{hom} : \overline{\boldsymbol{\varepsilon}}$$
$$\overline{\boldsymbol{\varepsilon}} = \mathbf{M}^{hom} : \overline{\boldsymbol{\sigma}} \quad (46)$$

where $\overline{\boldsymbol{\sigma}}$, $\overline{\boldsymbol{\varepsilon}}$ denote the average strain and stress tensors. It should be noted that the definition of the average strain is the same as the traditional one, while the definition of the average stress should consider the stress-jump across the interface [31]:

$$\overline{\boldsymbol{\varepsilon}} = (1-f)\overline{\boldsymbol{\varepsilon}}_m + f\overline{\boldsymbol{\varepsilon}}_i$$
$$\overline{\boldsymbol{\sigma}} = (1-f)\overline{\boldsymbol{\sigma}}_m + f\overline{\boldsymbol{\sigma}}_i + \frac{f}{V_i}\int_{V_i}(\overline{\boldsymbol{\sigma}}_m - \overline{\boldsymbol{\sigma}}_i)\cdot\mathbf{n}\otimes\mathbf{r}\,\mathrm{d}S \quad (47)$$

where $\overline{\boldsymbol{\varepsilon}}_j$ and $\overline{\boldsymbol{\sigma}}_j$ denote volume averages of the strain and stress in the matrix/

inhomogeneity. $f$ denotes the volume fraction of inhomogeneity. For a two-phase model, $\mathbf{L}^{\text{hom}}$ and $\mathbf{M}^{\text{hom}}$ are given in Duan et al. [31]:

$$\begin{aligned}\mathbf{L}^{\text{hom}} &= \mathbf{L}_m + f(\mathbf{L}_i - \mathbf{L}_m) : \mathbf{A}_i + f\mathbf{A}_\Gamma \\ \mathbf{M}^{\text{hom}} &= \mathbf{M}_m + f(\mathbf{M}_i - \mathbf{M}_m) : \mathbf{B}_i - f\mathbf{B}_\Gamma\end{aligned} \quad (48)$$

where $\mathbf{L}_j$ and $\mathbf{M}_j$ are the stiffness and compliance tensors of the two phases. $\mathbf{A}_i$, $\mathbf{B}_i$, $\mathbf{A}_\Gamma$ and $\mathbf{B}_\Gamma$ are defined as:

$$\begin{aligned}\bar{\boldsymbol{\varepsilon}}^i &= \mathbf{A}_i : \bar{\boldsymbol{\varepsilon}} \\ \bar{\boldsymbol{\sigma}}^i &= \mathbf{B}_i : \bar{\boldsymbol{\sigma}} \\ \frac{1}{V_i}\int_{V_i}(\bar{\boldsymbol{\sigma}}_m - \bar{\boldsymbol{\sigma}}_i) \cdot \mathbf{n} \otimes \mathbf{r} \, dS &= \mathbf{A}_\Gamma : \bar{\boldsymbol{\varepsilon}} \\ \frac{1}{V_i}\int_{V_i}(\bar{\boldsymbol{\sigma}}_m - \bar{\boldsymbol{\sigma}}_i) \cdot \mathbf{n} \otimes \mathbf{r} \, dS &= \mathbf{B}_\Gamma : \bar{\boldsymbol{\sigma}}\end{aligned} \quad (49)$$

Thus, once the concentration tensors $\mathbf{A}_i, \mathbf{B}_i$, $\mathbf{A}_\Gamma$ and $\mathbf{B}_\Gamma$ are determined, the effective stiffness tensor $\mathbf{L}^{\text{hom}}$ and the effective compliance tensor $\mathbf{M}^{\text{hom}}$ can be easily calculated.

The simplest method is the "dilute" approximation [46], in which $\mathbf{A}_i, \mathbf{B}_i$, $\mathbf{A}_\Gamma$ and $\mathbf{B}_\Gamma$ are approximated by considering the case of embedding a single nano-inhomogeneity in an all-matrix medium. Thus, we have

$$\begin{aligned}\mathbf{A}_i &= \bar{\mathbf{A}}^i \\ \mathbf{B}_i &= \bar{\mathbf{B}}^i \\ \mathbf{A}_\Gamma &= \bar{\mathbf{A}}^\Gamma \\ \mathbf{B}_\Gamma &= \bar{\mathbf{B}}^\Gamma\end{aligned} \quad (50)$$

where $\bar{\mathbf{A}}^i$, $\bar{\mathbf{B}}^i, \bar{\mathbf{A}}^\Gamma$ and $\bar{\mathbf{B}}^\Gamma$ are given in section 4. Substituting Eq.(50) into Eq.(48), we can express the effective stiffness tensor $\mathbf{L}^{\text{hom}}$ and compliance tensor $\mathbf{M}^{\text{hom}}$ as:

$$\begin{aligned}\mathbf{L}^{\text{hom}} &= \mathbf{L}_m + f(\mathbf{L}_i - \mathbf{L}_m) : \bar{\mathbf{A}}^i + f\bar{\mathbf{A}}^\Gamma \\ \mathbf{M}^{\text{hom}} &= \mathbf{M}_m + f(\mathbf{M}_i - \mathbf{M}_m) : \bar{\mathbf{B}}^i - f\bar{\mathbf{B}}^\Gamma\end{aligned} \quad (51)$$

Expressing $\mathbf{L}^{\text{hom}}$ as $3k_{\text{DA}}^{\text{hom}}\mathbf{J} + 2\mu_{\text{DA}}^{\text{hom}}\mathbf{K}$, we can obtain the effective bulk and shear modulus:

$$k_{DA}^{hom} = \frac{1}{3}\left(3k_m + \overline{A}_1^\Gamma f + f(3k_i - 3k_m)\overline{A}_1^i\right)$$
$$\mu_{DA}^{hom} = \frac{1}{2}\left(2\mu_m + \overline{A}_2^\Gamma f + f(2\mu_i - 2\mu_m)\overline{A}_2^i\right)$$
(52)

As discussed in[46], the dilute approximation neglects particle interactions and is therefore valid only for small volume fractions of reinforcements. In order to predict overall mechanical properties of nano-composites with higher volume fractions, we employ the Mori-Tanaka method (MTM) [47]. Following the procedure given in [48], we have the following relations:

$$\overline{\boldsymbol{\varepsilon}}^i = \overline{\mathbf{A}}^i : \overline{\boldsymbol{\varepsilon}}^m$$
$$\overline{\boldsymbol{\sigma}}^i = \overline{\mathbf{B}}^i : \overline{\boldsymbol{\sigma}}^m$$
$$\frac{1}{V_i}\int_{V_i}(\overline{\boldsymbol{\sigma}}_m - \overline{\boldsymbol{\sigma}}_i)\cdot \mathbf{n}\otimes \mathbf{r}\, dS = \overline{\mathbf{A}}^\Gamma : \overline{\boldsymbol{\varepsilon}}^m$$
$$\frac{1}{V_i}\int_{V_i}(\overline{\boldsymbol{\sigma}}_m - \overline{\boldsymbol{\sigma}}_i)\cdot \mathbf{n}\otimes \mathbf{r}\, dS = \overline{\mathbf{B}}^\Gamma : \overline{\boldsymbol{\sigma}}^m$$
(53)

Substituting Eq.(53) into Eq.(47), we can express $\mathbf{A}_i, \mathbf{B}_i,\ \mathbf{A}_\Gamma$ and $\mathbf{B}_\Gamma$ in terms of $\overline{\mathbf{A}}^i$, $\overline{\mathbf{B}}^i, \overline{\mathbf{A}}^\Gamma$ and $\overline{\mathbf{B}}^\Gamma$ as follows:

$$\mathbf{A}_i = \overline{\mathbf{A}}^i : \left[\mathbf{I}_2 + f(\overline{\mathbf{A}}^i - \mathbf{I}_2)\right]^{-1}$$
$$\mathbf{B}_i = \overline{\mathbf{B}}^i : \left[\mathbf{I}_2 + f(\overline{\mathbf{B}}^i + \overline{\mathbf{B}}^\Gamma - \mathbf{I}_2)\right]^{-1}$$
$$\mathbf{A}_\Gamma = \overline{\mathbf{A}}^\Gamma : \left[\mathbf{I}_2 + f(\overline{\mathbf{A}}^i - \mathbf{I}_2)\right]^{-1}$$
$$\mathbf{B}_\Gamma = \overline{\mathbf{B}}^\Gamma : \left[\mathbf{I}_2 + f(\overline{\mathbf{B}}^i + \overline{\mathbf{B}}^\Gamma - \mathbf{I}_2)\right]^{-1}$$
(54)

Substituting Eq.(54) into Eq.(48), $\mathbf{L}^{hom}$ and compliance tensor $\mathbf{M}^{hom}$ can be written as:

$$\mathbf{L}^{hom} = \mathbf{L}_m + f\left[(\mathbf{L}_i - \mathbf{L}_m):\overline{\mathbf{A}}^i + \overline{\mathbf{A}}^\Gamma\right]:\left[\mathbf{I}_2 + f(\overline{\mathbf{A}}^i - \mathbf{I}_2)\right]^{-1}$$
$$\mathbf{M}^{hom} = \mathbf{M}_m + f\left[(\mathbf{M}_i - \mathbf{M}_m):\overline{\mathbf{B}}^i - \mathbf{M}_m:\overline{\mathbf{B}}^\Gamma\right]:\left[\mathbf{I}_2 + f(\overline{\mathbf{B}}^i + \overline{\mathbf{B}}^\Gamma - \mathbf{I}_2)\right]^{-1}$$
(55)

And we can obtain the effective bulk and shear modulus:

$$k_{MTM}^{hom} = \frac{f\left(\overline{A}_1^\Gamma + 3k_i\overline{A}_1^i - 3k_m\right) + 3k_m}{3 + 3f\left(-1 + \overline{A}_1^i\right)}$$

$$\mu_{MTM}^{hom} = \frac{f\left(\overline{A}_2^\Gamma + 2\overline{A}_2^i\mu_i - 2\mu_m\right) + 2\mu_m}{2 + 2f\left(-1 + \overline{A}_2^i\right)}$$

(56)

Again, in addition to size-dependency, two interesting phenomena are observed from the closed form expressions of derived effective moduli. First, the effective shear moduli depend on the interface curvature parameter $\eta_s = 3\zeta_s + 5\chi_s$, that is to say, it is not affected individually by $\zeta_s$ and $\chi_s$ when $\eta_s$ is a fixed value. Second, the effective bulk modulus is invariant to interface bending resistance parameters. The invariability of effective bulk modulus with respect to interface bending resistance is unexpected since elastic fields in the composites are affected by $\eta_s$ as seen from Eqs.(A.33-A.47).

### 5.2. Numerical results and discussion

In this subsection, we conduct a series of parametric studies to investigate the influence of surface bending resistance on the effective shear modulus. Material properties for the inhomogeneity are $E_i = 410$ GPa and $v_i = 0.14$, while material properties for the matrix are $E_m = 71$ GPa and $v_m = 0.35$. The interface parameters are selected as $\lambda_s = 3.4939$ N/m and $\mu_s = -5.4251$ N/m [49]. The radius of the inhomogeneity is $R = 1$nm.

Fig.2 and Fig.3 shows the effective shear modulus calculated by different methods with different interface bending stiffness parameters and with different volume fractions of the inhomogeneity. Results reveal that the interface bending stiffness parameters and volume fractions can significantly affect the effective shear modulus. From Fig.2, it is also observed that the effective modulus calculated by the dilute approximation agrees with that obtained by the Mori-Tanaka method when the volume fraction of inhomogeneities is small, but they differ considerably when the volume fraction is large, because the dilute approximation neglects any interaction between inhomogeneities. Therefore, we use the effective modulus calculated by the Mori-Tanaka method hereinafter. Besides, the effective shear modulus computed by Eq.(56) when $\eta_s = 0$ agrees with the effective shear modulus using the G-M interface model[30], which partially verifies the effective moduli as derived in this paper.

An interesting phenomenon is observed in Fig.3 that the effective modulus is significantly affected by interface bending when $\eta_s$ is near the characteristic curvature parameter $\eta_s^*$ ($\eta_s^*$ is given in the Appendix). It should be pointed out that the effective modulus calculated by the Mori-Tanaka method is expected to have larger errors near $\eta_s^*$, because it is the singular point of Eq.(56). However, it is interesting to learn that such a singular point exists near which the effective shear modulus can be significantly increased by surface modification. We further study the influences of the radius and the volume fraction of the inhomogeneity on $\eta_s^*$. It can be seen in Fig.4 that $\eta_s^*$ is significantly increased with the increase of the inhomogeneity's size. In contrast, volume fraction has little effect on $\eta_s^*$.

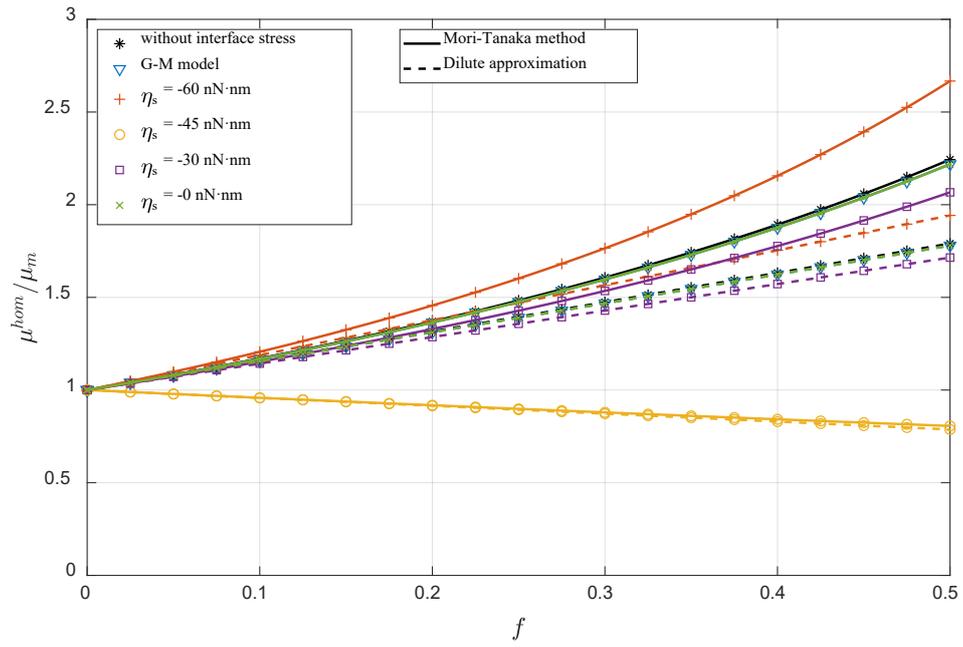

Fig.2. Variation of the effective shear modulus with different volume fractions using two homogenization methods

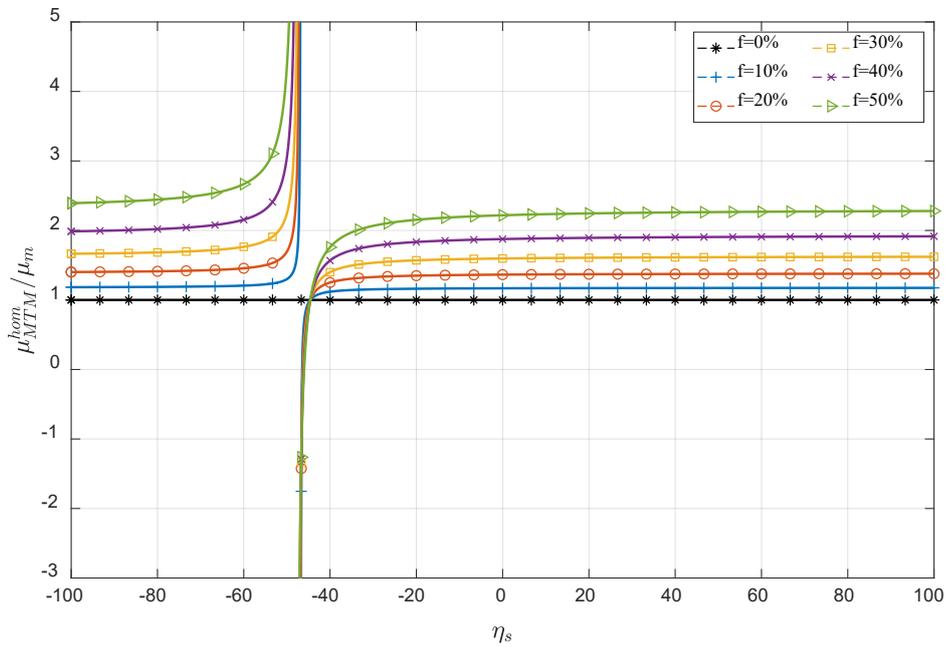

Fig.3. Influence of the interface curvature parameter $\eta_s$ on the effective shear modulus

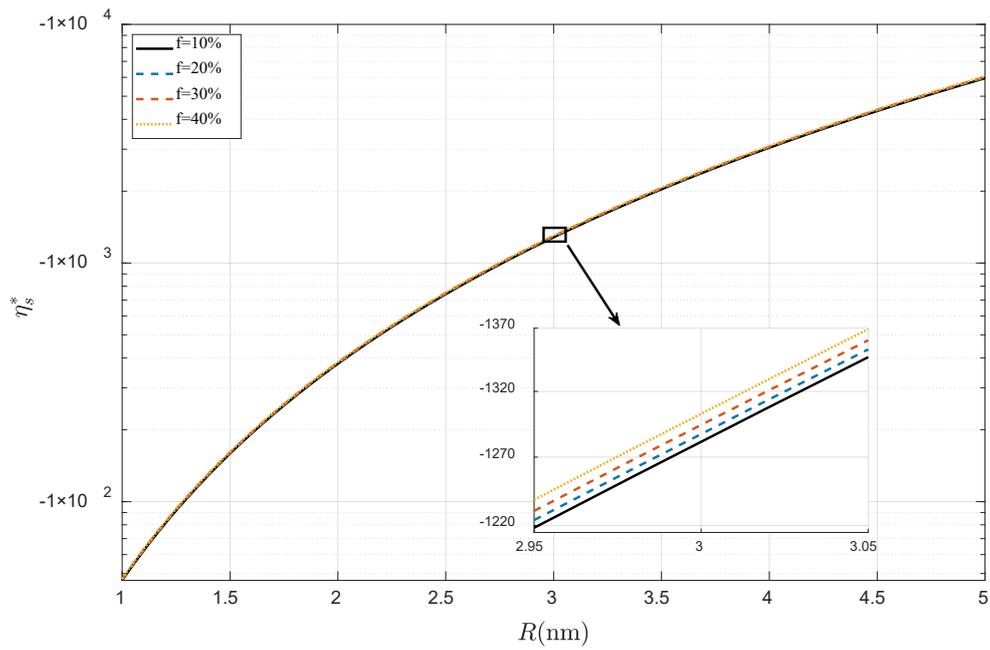

Fig.4. The variation of $\eta_s^*$ with different inhomogeneity radius and different volume fractions

A nano-porous material with the S-O model is also investigated in this study. The material properties for the matrix are $E_m = 71$ GPa and $v_m = 0.35$. We choose two sets of interface elastic constants: (a) $\lambda_s = 3.4939$ N/m, $\mu_s = -5.4251$ N/m and (b) $\lambda_s = 6.8511$ N/m, $\mu_s = -0.376$ N/m [49]. The variation of the effective shear modulus with the interface curvature parameter and volume fractions is illustrated in Fig.5, when the radius of the pore is $R = 1$nm. Fig.6 shows the variation of the effective shear modulus of the porous material with the pore's radius. Size-dependency is observed. The smaller the void is, the more significant the interface effects are. Results also reveal that the effective shear moduli of Nano-porous materials can be greatly improved by an appropriate surface modification.

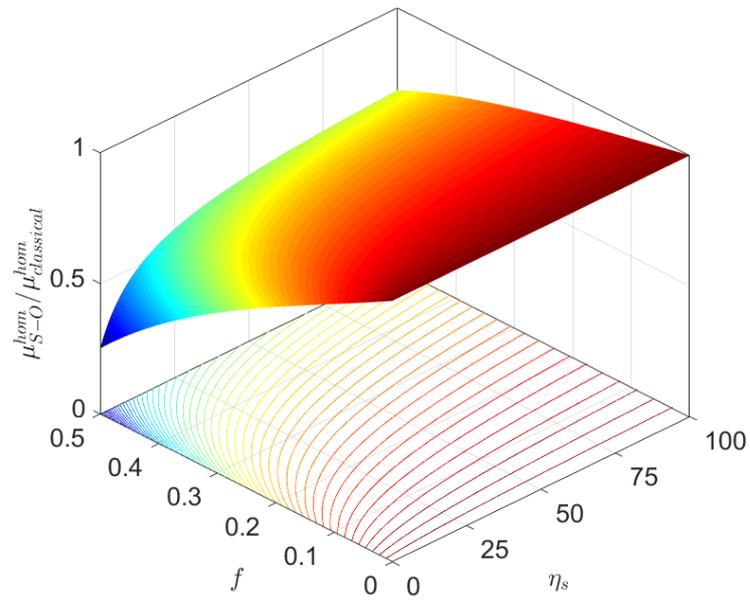

(a)

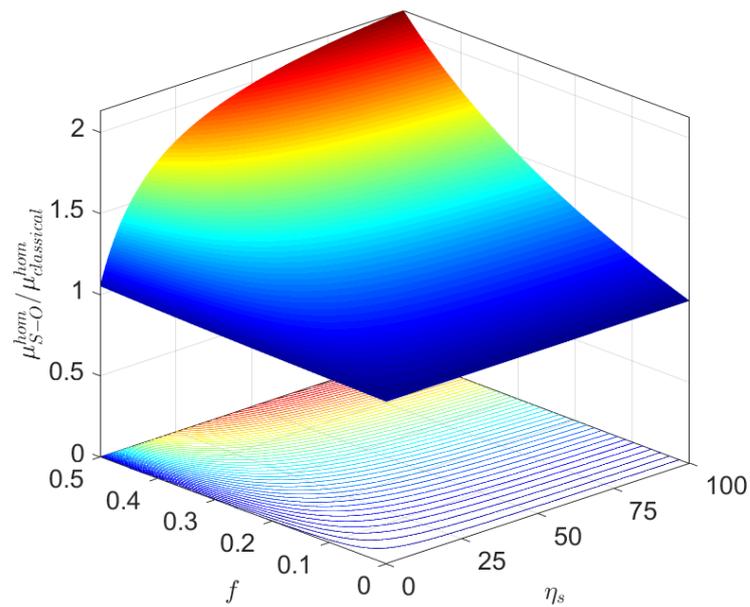

(b)

Fig.5. The variation of the effective shear modulus with different surface curvature constants and volume fractions when (a) $\lambda_s = 3.4939$ N/m, $\mu_s = -5.4251$ N/m and (b) $\lambda_s = 6.8511$ N/m, $\mu_s = -0.376$ N/m

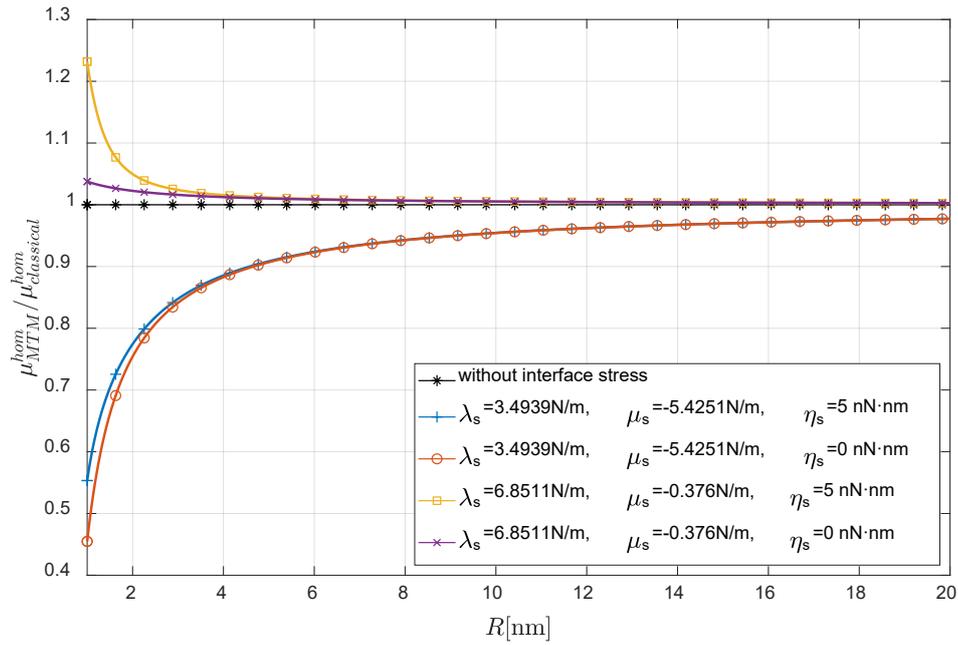

Fig.6. The influence of the void's radius on the effective shear modulus

## 6. Conclusions

In this study, we generalized the fundamental framework of analytical micromechanics to study nano-composites with both interface stretching and bending effects.

The first contribution of this paper is that the explicit analytical solutions for spherical nano-inclusion with an interface defined by the Steigmann–Ogden model is derived, considering uniform eigenstrains or far-field stress/strain loadings. The Eshelby tensors and the stress/strain concentration tensors for these problems are also derived in this paper for the first time.

The second contribution of this paper is that explicit expressions for the effective bulk/shear moduli for nano-composites with a Steigmann–Ogden(S–O) interface are provided. In addition to size-dependency, the closed form expressions show that the effective bulk modulus is invariant to interface bending resistance parameters, in contrast to the effective shear modulus.

The third contribution of this paper is that an equivalent interface curvature parameter $\eta_s$ and a characteristic value $\eta_s^*$ for this curvature parameter are defined. The concentration tensors and the effective shear modulus are not affected individually by $\zeta_s$ and $\chi_s$, when $\eta_s$ is a fixed value. The effective shear modulus is affected significantly by the interface bending resistance when $\eta_s$ is near $\eta_s^*$.

The explicit solutions derived in this paper can be used as a benchmark for semi-analytical solutions and numerical solutions in future studies. The derived explicit expression of the effective modulus with the Steigmann–Ogden interface model is provided in the supplemental MATLAB code for the convenience of users.

**Acknowledgement**

The first three authors thankfully acknowledge the support from the National Key Research and Development Program of China (No. 2017YFA0207800), the National Natural Science Foundation of China (grant No. 11872008), and the Fundamental Research Funds for the Central Universities.

**Appendix**

For the problem of a nano-inclusion, coefficients in Eqs.(9-21) are:

$$\eta_s = 3\zeta_s + 5\chi_s \tag{A.1}$$

$$\begin{aligned}\gamma_s = &(2\eta_s(8(-4+5v_m)(-7+10v_m)\lambda_s + 15R(-1+v_m)(-28+39v_m)\mu_m \\ &+20(-4+5v_m)(-7+10v_m)\mu_s) + R^2(525R^2(-1+v_m)^2\mu_m^2 + 8(-4+5v_m \\ &)(-7+10v_m)\mu_s(\lambda_s+\mu_s) + 10R(-1+v_m)\mu_m(-56\lambda_s + 79v_m\lambda_s - 98\mu_s + \\ &133v_m\mu_s)));\end{aligned} \tag{A.2}$$

$$M_1 = -\frac{R^4(1+v_m)\mu_m}{6(-1+2v_m)\lambda_s + 9R(-1+v_m)\mu_m + 6(-1+2v_m)\mu_s}; \tag{A.3}$$

$$M_2 = (R^4\mu_m(32(7-10v_m)\eta_s + R^2(24(7-10v_m)\lambda_s - 175R(-1+v_m)\mu_m \\ + 28(7-10v_m)\mu_s)))/(6\gamma_s); \tag{A.4}$$

$$M_3 = (R^6\mu_m(16(-2+v_m)(-7+10v_m)\eta_s + R^2(-8(1+v_m)(-7+10v_m) \\ \lambda_s - 105R(-1+v_m)\mu_m + 28\mu_s + 8(9-20v_m)v_m\mu_s)))/(3\gamma_s); \tag{A.5}$$

$$C_1 = -\frac{R(1+v_m)\mu_m}{6(-1+2v_m)\lambda_s + 9R(-1+v_m)\mu_m + 6(-1+2v_m)\mu_s}; \tag{A.6}$$

$$C_2 = (-4(-4+5v_m)\mu_m(2\eta_s - R^2(\lambda_s + 2\mu_s)))/(3R\gamma_s); \tag{A.7}$$

$$C_3 = (2R(-4+5v_m)\mu_m(4(-7+16v_m)\eta_s + R^2(6(-7+8v_m)\lambda_s + 7(-1+ \\ v_m)(5R\mu_m + 8\mu_s))))/(3\gamma_s); \tag{A.8}$$

$$M_{1\alpha\beta} = \delta_{\alpha\beta} M_1 \varepsilon^*_{\alpha\beta} \tag{A.9}$$

$$M_{2\alpha\beta} = M_2 \varepsilon^*_{\alpha\beta} \tag{A.10}$$

$$M_{3\alpha\beta} = M_3 \varepsilon^*_{\alpha\beta} \tag{A.11}$$

$$C_{1\alpha\beta} = \delta_{\alpha\beta} C_1 \varepsilon^*_{\alpha\beta} \tag{A.12}$$

$$C_{2\alpha\beta} = C_2 \varepsilon^*_{\alpha\beta} \tag{A.13}$$

$$C_{3\alpha\beta} = C_3 \varepsilon^*_{\alpha\beta} \tag{A.14}$$

where $\alpha, \beta = x, y, z$ and $\delta$ is Kronecker delta.

For an inhomogeneity under far-field stresses, coefficients in Eqs.(24-37) are:

$$\gamma_\Sigma = 2\eta_s(8(-7+10v_i)(-4+5v_m)\lambda_s + R(-49+61v_i)(-4+5v_m)\mu_i + 4R \\ (-7+10v_i)(-8+7v_m)\mu_m + 20(-7+10v_i)(-4+5v_m)\mu_s) + R^2(-R^2(7 \\ \mu_i + 5v_i\mu_i + 28\mu_m - 40v_i\mu_m)(-8\mu_i + 10v_m\mu_i - 7\mu_m + 5v_m\mu_m) + 8(-7+ \\ 10v_i)(-4+5v_m)\mu_s(\lambda_s + \mu_s) + 2R((-35+47v_i)(-4+5v_m)\lambda_s\mu_i + 4(-7+ \\ 10v_i)(-5+4v_m)\lambda_s\mu_m + 7(7(-1+v_i)(-4+5v_m)\mu_i + 6(-7+10v_i)(-1+ \\ v_m)\mu_m)\mu_s)); \tag{A.15}$$

$$M_1^\Sigma = \frac{1-2v_m}{6\mu_m + 6v_m\mu_m}; \tag{A.16}$$

$$M_2^\Sigma = (R^3(2(-1+2v_i)(-1+2v_m)\lambda_s - R(1+v_i)(-1+2v_m)\mu_i + R(-1+2v_i)(1+v_m)\mu_m + 2(-1+2v_i)(-1+2v_m)\mu_s))/(6(1+v_m)\mu_m((-2+4v_i)\lambda_s - R(\mu_i + v_i\mu_i + 2\mu_m - 4v_i\mu_m) + 2(-1+2v_i)\mu_s));$$
(A.17)

$$M_3^\Sigma = \frac{1}{6\mu_m};$$
(A.18)

$$M_4^\Sigma = -5R^3(R^4(\mu_i - \mu_m)((7+5v_i)\mu_i + 4(7-10v_i)\mu_m) + R\eta_s((49-61v_i)\mu_i + 4(-7+10v_i)\mu_m) - 4R^2(-7+10v_i)\mu_s(\lambda_s + \mu_s) - 4(-7+10v_i)\eta_s(2\lambda_s + 5\mu_s) + R^3((35-47v_i)\lambda_s\mu_i + 4(-7+10v_i)\lambda_s\mu_m - 49(-1+v_i)\mu_i\mu_s))/(12\mu_m\gamma_\Sigma));$$
(A.19)

$$M_5^\Sigma = \frac{6}{5}M_4^\Sigma R^2 + \frac{-2R^6}{\gamma_\Sigma}(-7+10v_i)(-1+v_m)(-6\zeta_s + R^2\lambda_s + 2R^2\mu_s - 10\chi_s);$$
(A.20)

$$C_1^\Sigma = \frac{R(-1+2v_i)(-1+v_m)}{2(1+v_m)((2-4v_i)\lambda_s + R(\mu_i + v_i\mu_i + 2\mu_m - 4v_i\mu_m) + 2\mu_s - 4v_i\mu_s)};$$
(A.21)

$$C_2^\Sigma = (5(-1+v_m)(2\eta_s - R^2(\lambda_s + 2\mu_s)))/(-R\gamma_\Sigma);$$
(A.22)

$$C_3^\Sigma = -5R(-1+v_m)((28-64v_i)\eta_s + R^2(6(7-8v_i)\lambda_s + R(7\mu_i + 5v_i\mu_i + 28\mu_m - 40v_i\mu_m) - 56(-1+v_i)\mu_s))/(2\gamma_\Sigma));$$
(A.23)

$$M_{1\alpha\beta}^\Sigma = \delta_{\alpha\beta}M_1^\Sigma\Sigma_{\alpha\beta}$$
(A.24)

$$M_{2\alpha\beta}^\Sigma = \delta_{\alpha\beta}M_2^\Sigma\Sigma_{\alpha\beta}$$
(A.25)

$$M_{3\alpha\beta}^\Sigma = M_3^\Sigma\Sigma_{\alpha\beta}$$
(A.26)

$$M_{4\alpha\beta}^\Sigma = M_4^\Sigma\Sigma_{\alpha\beta}$$
(A.27)

$$M_{5\alpha\beta}^\Sigma = M_5^\Sigma\Sigma_{\alpha\beta}$$
(A.28)

$$C_{1\alpha\beta}^\Sigma = \delta_{\alpha\beta}C_1^\Sigma\Sigma_{\alpha\beta}$$
(A.29)

$$C_{2\alpha\beta}^\Sigma = C_2^\Sigma\Sigma_{\alpha\beta}$$
(A.30)

$$C_{3\alpha\beta}^\Sigma = C_3^\Sigma\Sigma_{\alpha\beta}$$
(A.31)

$$\eta_s^* = (R^2(R^2((7+5v_i)\mu_i + 4(7-10v_i)\mu_m)(2(-1+f)(-4+5v_m)\mu_i + (7 \\ +8f-5(1+2f)v_m)\mu_m) - 8(-1+f)(-7+10v_i)(-4+5v_m)\mu_s(\lambda_s+\mu_s) \\ +R(-2(-1+f)(-35+47v_i)(-4+5v_m)\lambda_s\mu_i + 8(-7+10v_i)(-5+4v_m+ \\ f(-4+5v_m))\lambda_s\mu_m - 98(-1+f)(-1+v_i)(-4+5v_m)\mu_i\mu_s + 84(-7+10 \\ v_i)(-1+v_m)\mu_m\mu_s)))/(2(8(-1+f)(-7+10v_i)(-4+5v_m)\lambda_s+(-1+f) \\ R(-49+61v_i)(-4+5v_m)\mu_i - 4R(-7+10v_i)(-8-4f+7v_m+5fv_m)\mu_m \\ +20(-1+f)(-7+10v_i)(-4+5v_m)\mu_s)) \quad (A.32)$$

where $\alpha, \beta = x, y, z$ and $\delta$ is Kronecker delta.

For an inhomogeneity with far-field strains, coefficients in Eqs.(39-45) are:

$$M_1^E = M_3^E = \frac{1}{3}; \quad (A.33)$$

$$M_2^E = -\frac{2(1+v_m)\mu_m}{-1+2v_m}M_2^\Sigma; \quad (A.34)$$

$$M_4^E = 2\mu_m M_4^\Sigma; \quad (A.35)$$

$$M_5^E = 2\mu_m M_5^\Sigma; \quad (A.36)$$

$$C_1^E = -\frac{2(1+v_m)\mu_m}{-1+2v_m}C_1^\Sigma; \quad (A.37)$$

$$C_2^E = 2\mu_m C_2^\Sigma; \quad (A.38)$$

$$C_3^E = 2\mu_m C_3^\Sigma; \quad (A.39)$$

$$M_{1\alpha\beta}^E = \delta_{\alpha\beta} M_1^E E_{\alpha\beta} \quad (A.40)$$

$$M_{2\alpha\beta}^E = \delta_{\alpha\beta} M_2^E E_{\alpha\beta} \quad (A.41)$$

$$M_{3\alpha\beta}^E = M_3^E E_{\alpha\beta} \quad (A.42)$$

$$M_{4\alpha\beta}^E = M_4^E E_{\alpha\beta} \quad (A.43)$$

$$M_{5\alpha\beta}^E = M_5^E E_{\alpha\beta} \quad (A.44)$$

$$C_{1\alpha\beta}^E = \delta_{\alpha\beta} C_1^E E_{\alpha\beta} \quad (A.45)$$

$$C_{2\alpha\beta}^E = C_2^E E_{\alpha\beta} \quad (A.46)$$

$$C_{3\alpha\beta}^E = C_3^E E_{\alpha\beta} \quad (A.47)$$